\documentclass{article}
\topskip=\baselineskip
\let\a=\alpha

\let\e=\emph

\let\ct=\cite
\let\lf=\left
\let\rt=\right
\let\bv=\mathbf
\let\dt=\cdot
\let\del=\nabla

\let\q=\widehat
\let\Q=\overbrace
\let\h=\hbar
\let\rta=\rightarrow
\let\dy=\displaystyle

\let\hl=\hfill
\newcommand{\m}{\mbox}
\newcommand{\ol}[1]{\makebox[\textwidth][s]{#1}}
\newcommand{\id}{\mathrm{I}}

\newcommand{\hf}{\ensuremath{{\scriptstyle\frac{1}{2}}}}

\newcommand{\be}{\begin{equation}}
\newcommand{\ee}{\end{equation}}
\newcommand{\dd}[3]{\\ \m{}\\ \ol{\m{#1}\hl\m{${\dy #2}$}\hl\m{#3}}\\ \m{}\\}
\newcommand{\re}[2]{\dd{}{#1}{(#2)}}

\newcommand{\de}[1]{\dd{}{#1}{}}
\newcommand{\ba}{\begin{array}}
\newcommand{\ea}{\end{array}}
\newcommand{\bea}{\begin{eqnarray}}
\newcommand{\eea}{\end{eqnarray}}
\newcommand{\beas}{\begin{eqnarray*}}
\newcommand{\eeas}{\end{eqnarray*}}

\newcommand{\vq}{\bv{q}}

\newcommand{\vp}{\bv{p}}

\newcommand{\vk}{\bv{k}}
\newcommand{\vl}{\bv{l}}
\newcommand{\qvq}{\q{\vq}}
\newcommand{\qvp}{\q{\vp}}
\newcommand{\qq}{\q{q}}
\newcommand{\qp}{\q{p}}
           \title{Unambiguous quantization from \\
  the maximum classical correspondence that is self-consistent: \\
  the slightly stronger canonical commutation rule Dirac missed}
\author{Steven Kenneth Kauffmann \\
        American Physical Society Senior Life Member}
\date{43 Bedok Road \\
      \# 01-11 \\
      Country Park Condominium \\
      Singapore 469564 \\
      Tel \& FAX: +65 6243 6334 \\
      Handphone: +65 9370 6583 \\
      \m{} \\
      and \\
      \m{} \\
      Unit 802, Reflection on the Sea \\
      120 Marine Parade \\
      Coolangatta QLD 4225 \\
      Australia \\
      Tel/FAX: +61 7 5536 7235 \\
      Mobile:  +61 4 0567 9058 \\
      \m{} \\
      Email: SKKauffmann@gmail.com}
\begin{document}
\maketitle
\begin{abstract}
Dirac's identification of the quantum analog of the Poisson bracket with the
commutator is reviewed, as is the threat of self-inconsistent overdetermination
of the quantization of classical dynamical variables which drove him to restrict
the assumption of correspondence between quantum and classical Poisson brackets
to embrace only the Cartesian components of the phase space vector.  Dirac's
canonical commutation rule fails to determine the order of noncommuting factors
within quantized classical dynamical variables, but does imply the
quantum/classical correspondence of Poisson brackets between any linear function
of phase space and the sum of an arbitrary function of only configuration space
with one of only momentum space.  Since every linear function of phase space is
itself such a sum, it is worth checking whether the assumption of
quantum/classical correspondence of Poisson brackets for all such sums is still
self-consistent.  Not only is that so, but this slightly stronger canonical
commutation rule also unambiguously determines the order of noncommuting
factors within quantized dynamical variables in accord with the 1925
Born-Jordan quantization surmise, thus replicating the results of the
Hamiltonian path integral, a fact first realized by E. H. Kerner.  Born-Jordan
quantization validates the generalized Ehrenfest theorem, but has no inverse,
which disallows the disturbing features of the poorly physically motivated
invertible Weyl quantization, i.e., its unique deterministic classical ``shadow
world'' which can manifest negative densities in phase space.
\end{abstract}

\subsection*{Introduction}

The canonical commutation rule and the Heisenberg equation of motion
both give concrete expression to Dirac's profound discovery that
$(-i/\h)$ times the commutator bracket is the quantum analog of the
classical Poisson bracket, and together serve to incorporate both the
correspondence principle and the uncertainty principle into orthodox
operator quantum dynamics.  Dirac's 1925 version of the canonical
commutation rule is well-known, however, to be too weak to determine
the ordering of noncommuting factors that in principle can occur in
the quantization of an arbitrary classical dynamical variable---albeit
this has never been a significant issue in practice because such
factors rarely feature in the classical Hamiltonians that are
hypothesized for those physical systems for which quantum dynamics
appears to be useful.  As a matter of principle, however, this
ordering ambiguity in the quantization of general classical dynamical
variables can obviously be viewed as an annoying gap in the
theoretical completeness of orthodox quantum dynamics.  Dirac in 1925
was under the impression that he had little choice but to opt for his
weak version of the canonical commutation rule because the most
obvious stronger alternative turns out to self-inconsistently
\e{overdetermine} the quantization of general classical dynamical
variables, and obviously an annoying apparent gap in theoretical
completeness is a lesser evil than outright self-inconsistency.  The
fleshing out of the alternative Hamiltonian phase-space path integral
approach to quantum dynamics in the late 1960's, however, yielded an
unambiguous quantization of \e{all} classical dynamical variables---a
groundbreaking result which unfortunately was soon mistakenly disputed.
That result motivated reexamination of the range of possible canonical
commutation rules, which led to the realization that a \e{slightly
stronger one} than Dirac's weak 1925 version \e{still retains the
latter's self-consistency} but nevertheless \e{completely resolves
its ordering ambiguity}---this slightly stronger canonical commutation
rule in fact yields \e{exactly the same} unambiguous quantization of
all classical dynamical variables as that which is implied by the
Hamiltonian phase-space path integral.  Lamentably, as a
psychologically freighted consequence of the erroneous disputation of
the unambiguous quantization result for the Hamiltonian phase-space
path integral, this slightly stronger canonical commutation rule was
never published nor publicly disclosed by its original discoverer.
That notwithstanding, the fact is that Dirac in 1925 \e{was intimately
familiar with all the knowledge and tools that are needed for its
discovery}; it is a matter of mere \e{historical happenstance} that he
failed to light on it \e{at that time}.  Thus it was \e{very far from
inevitable} that the ordering ambiguity gap in the theoretical
completeness of quantum dynamics \e{should even have occurred}.

\subsection*{Dirac's quantum analog of the classical Poisson bracket}

By way of placing Heisenberg's matrix quantum mechanics on a more
general footing, Dirac~\ct{Di}
abstracted Heisenberg's \e{matrix} quantum dynamical variables as
simply \e{noncommuting} quantum dynamical variables of the form
$\q{F}(\qvq ,\qvp )$, whose underlying phase space vector
$(\qvq ,\qvp )$ \e{also} consists of mutually noncommuting compo%
nents.  In step with Heisenberg's practice, Dirac required these
quantized dynamical variables to obey equations of motion which
correspond as \e{closely} to the \e{classical Hamiltonian} equa%
tions of motion that are satisfied by their \e{unquantized prede%
cessors} as their \e{noncommuting character} can accommodate.  It
was therefore envisaged that the classical equation of motion,
\re{
dF/dt = \{ F,H\} ,
}{1a}
where $\{\, ,\} $ is the Poisson bracket, has the quantized counterpart,
\re{
d\q{F}/dt = \{ \q{F},\q{H}\}_Q ,
}{1b}
where $\{\, ,\}_Q $ is Dirac's \e{quantum analog} of the Poisson bracket.
In parallel with $\{\, ,\} $, $\{\, ,\}_Q $ is assumed to be \e{bilinear}
in its two arguments (its linearity in its \e{first} argument is already
implied by the linearity of the time derivative operation on the left
hand side of Eq.~(1b)).  Since the classical Hamiltonian $H$ is a con%
stant of motion, its quantization $\q{H}$ is postulated to be so as well,
i.e., $d\q{H}/dt=0$, which, from Eq.~(1b), implies that
$\{ \q{H},\q{H}\}_Q=0$.  Given two quantized Hamiltonians, $\q{H_1}$ and
$\q{H_2}$, their sum $\q{H}=\q{H_1}+\q{H_2}$ is also a quantized Hamil%
tonian.  This, together with the \e{bilinearity} of $\{\, ,\}_Q$ and the
\e{vanishing} of $\{ \q{H},\q{H}\}_Q$, $\{ \q{H_1},\q{H_1}\}_Q$, and
$\{ \q{H_2},\q{H_2}\}_Q$, obviously implies that,
\re{
\{ \q{H_1},\q{H_2}\}_Q + \{ \q{H_2},\q{H_1}\}_Q = 0 .
}{2a}
Now the only evident distinction between a quantized Hamiltonian
$\q H(\qvq ,\qvp )$ and a general quantized dynamical variable
$\q F(\qvq ,\qvp )$ is that $\q H(\qvq ,\qvp )$ has dimensions of
energy; therefore multiplying the arbitrary $\q F(\qvq ,\qvp )$ by
a nonzero \e{constant} (in the quantized phase space variariables
$(\qvq ,\qvp )$) that has the appropriate \e{dimensions} will \e{change}
it to such a quantized $\q H(\qvq ,\qvp )$.  This fact, together with
the \e{bilinearity} of $\{\, ,\}_Q$ and Eq.~(2a), permits us to conclude
that $\{\, ,\}_Q$, like $\{\, ,\}$, is always \e{antisymmetric} in its
two arguments, i.e.,
\re{
\{ \q{F_1},\q{F_2}\}_Q = - \{ \q{F_2},\q{F_1}\}_Q .
}{2b}
\indent
The time derivative of the \e{product} of two noncommuting quantized
dynamical variables $\q F$ and $\q G$ (e.g., Heisenberg's matrices) is
well-known to be given by the familiar \e{product differentiation rule},
but with the \e{order} of the two factors always \e{strictly maintained},
\re{
d(\q F\q G)/dt = (d\q F/dt)\q G + \q F(d\q G /dt).
}{3a}
This, together with Eq.~(1b), implies that,
\re{
\{ \q F \q G ,\q H \}_Q = \{ \q F ,\q H \}_Q \q G + \q F \{ \q G ,\q H \}_Q,
}{3b}
and we can, upon taking the \e{bilinearity} of $\{\, ,\}_Q$ into account
along the lines discussed just before Eq.~(2b), substitute an \e{arbitrary}
quantized dynamical variable for the $\q H$ which appears in Eq.~(3b).
Of course the \e{order} of the factors that occur in Eq.~(3b) must be
strictly maintained---the classical Poisson bracket $\{\, ,\}$ satisfies
a relation that formally parallels Eq.~(3b), but there the ordering of
the factors that occur makes no difference.  Given four arbitrary quan%
tized dynamical variables $\q{F_1}$, $\q{G_1}$, $\q{F_2}$, $\q{G_2}$,
Dirac started with the relation,
\de{
\{\q{F_1}\q{G_1},\q{F_2}\q{G_2}\}_Q = - \{\q{F_2}\q{G_2},\q{F_1}\q{G_1}\}_Q ,
}
which follows from the antisymmetry of $\{\, ,\}_Q$ (given by Eq.~(2b)).
He then systematically applied Eqs.~(3b) and (2b) in repeated succession
to both its left and right hand sides, always keeping scrupulous track of
the \e{order} of multiplicative factors.  After taking account of the
cancellation of identical terms, there results,
\re{
(\q{F_1}\q{F_2}-\q{F_2}\q{F_1})\{\q{G_1},\q{G_2}\}_Q =
\{\q{F_1},\q{F_2}\}_Q(\q{G_1}\q{G_2}-\q{G_2}\q{G_1}) .
}{4}
If $\q{F_1}$ commuted with $\q{F_2}$ and $\q{G_1}$ with $\q{G_2}$, this
would be an implication-free \e{identity}---as its \e{classical} Poisson
bracket analog indeed \e{is}.  For the \e{arbitrary} noncommuting
\e{quantized} dynamical variables $\q{F_1}$, $\q{G_1}$, $\q{F_2}$,
and $\q{G_2}$, however, Eq.~(4) implies that,
\re{ \ba {rcl}
\{\q{G_1},\q{G_2}\}_Q &  =  & K(\q{G_1}\q{G_2}-\q{G_2}\q{G_1})\;\; \m{and} \\
\{\q{F_1},\q{F_2}\}_Q &  =  & K(\q{F_1}\q{F_2}-\q{F_2}\q{F_1}) , \\
                      &     &
\ea }{5a}
where $K$ is a \e{universal} constant.  By referring to the equation of
motion of Heisenberg's Hermitian matrix quantum dynamical variables, Dirac
could see that $K$ was equal to the imaginary universal constant $-i/\h $.
Therefore, $\{\, ,\}_Q$ is given by the \e{commutator} expression,
\re{
\{\q{F_1},\q{F_2}\}_Q = (-i/\h )(\q{F_1}\q{F_2}-\q{F_2}\q{F_1}) =
(-i/\h )[\q{F_1} ,\q{F_2} ] .
}{5b}
\mbox{}

\subsection*{Exploring potential canonical commutation rules}

With the definitive commutator expression of Eq.~(5b) in hand for
$\{\, ,\}_Q$, Dirac was naturally tempted to postulate the following
complete correspondence of the quantum Poisson bracket $\{\, ,\}_Q$
to its classical counterpart $\{\, ,\}$,
\re{
\{\q{F_1},\q{F_2}\}_Q = \Q{ \{ F_1, F_2\} }\;\, ,
}{6}
where we have used the overbrace symbol synonymously with the hat
symbol to denote the quantization of a classical dynamical variable
(we use the overbrace mainly where lack of sufficient extensibility
of the hat symbol presents a problem).  For a dynamical system
having one degree of freedom, Eq.~(6) implies that $\Q{qp} =
\hf (\qq\qp + \qp\qq )$, $\Q{q^2p} = \qq\qp\qq$, and $\Q{qp^2} =
\qp\qq\qp$.  However, for the still higher order quantization
$\Q{q^2p^2}$, the \e{two} relations
$\{\:\Q{q^2p}\: ,\:\Q{qp^2}\:\}_Q = \Q{ \{ q^2p, qp^2\} }$ and
$\{\,\Q{q^3}\, ,\,\Q{p^3}\,\}_Q = \Q{ \{ q^3, p^3\} }$ that are
\e{both} implied by Eq.~(6) produce results which unfortunately
\e{differ} from each other by $\h^2/3$.  Though he did not publish
the details (the calculations just sketched were first published
by Groenewold~\ct{Gr}
many years later), Dirac was acutely aware that the straight%
forward postulate of Eq.~(6) self-inconsistently \e{overdeter%
mines} the quantization of classical dynamical variables.  Casting
about for a way to avoid such overdetermination, Dirac \e{restricted}
the classical dynamical variables $F_1$ and $F_2$ that appear in
Eq.~(6) to be \e{only} the Cartesian components of the phase space
vector $(\vq, \vp )$.  For these, the right hand side of Eq.~(6) is
invariably the quantization of either zero or unity, which are
naturally taken to be, respectively, zero and the quantum identity
$\id$ (which is, for example, the identity matrix in the Heisenberg
matrix quantum mechanics).  Therefore, Dirac's restriction
\e{prevents} Eq.~(6) from \e{determining} the quantization of
\e{any} nonconstant classical dynamical variable, which obviously
(if hardly elegantly!) \e{eliminates} the issue of its possible
\e{overdetermination}.  The obvious downside of Dirac's drastic
restriction on Eq.~(6) is exactly this resulting \e{nondetermina%
tion} of the quantizations of nonconstant classical dynamical
variables.  Even the \e{additional} requirements that such quanti%
zations be \e{Hermitian} matrices in the Heisenberg matrix quantum
mechanics (for \e{real} classical dynamical variables), and that
they reduce properly in the limit that the Cartesian components of
$(\qvq ,\qvp )$ all \e{commute} with each other (i.e., when
$\h\rta 0$), still leaves \e{major} ambiguities: such a simple
entity as $\Q{qp}$ cannot be determined beyond $(\hf - ik)\qq\qp +
(\hf + ik)\qp\qq = \hf (\qq\qp + \qp\qq ) + k\h$, where k is an
\e{arbitrary} real constant.  The main reason that Dirac's drastic
restriction on Eq.~(6) has been \e{tolerated} for so long is
the prevalence in practice of classical Hamiltonians of the special
form $|\vp |^2/(2m) + V(\vq )$, whose standard quantization as
$|\qvp |^2/(2m) + V(\qvq )$ follows from supplementing Dirac's
restricted form of Eq.~(6) with the compatible further implicit
natural assumptions that $\Q{F(\vp )} = F(\qvp )$, that $\Q{G(\vq )}
= G(\qvq )$, and that quantization is a \e{linear} process. These
implicit assumptions, plus the additional one that $\q{c} = c\,\id $
for any $c$ that is \e{constant} in $(\vq ,\vp )$, are needed
auxiliaries to postulation of Dirac's restricted form of Eq.~(6).
Taking account of these implicit auxiliary assumptions and of
the bilinearity of $\{\, ,\}_Q$ and $\{\, ,\}$, Dirac's
restricted form of Eq.~(6) can be extended and reexpressed as,
\re{ \ba {c}
\{\;\Q{c_1 + \vk_1\dt\vq + \vl_1\dt\vp}\; ,
\;\Q{c_2 + \vk_2\dt\vq + \vl_2\dt\vp}\;\}_Q
   =   \Q{ \{ c_1 + \vk_1\dt\vq + \vl_1\dt\vp, c_2 + \vk_2\dt\vq
       + \vl_2\dt\vp\} } \\
 \m{}                     \\
   =   (\vk_1\dt\vl_2 - \vl_1\dt\vk_2 )\,\id , \\
 \m{}
 \ea }{7}
where $c_1$, $c_2$, $\vk_1$, $\vk_2$, $\vl_1$, and $\vl_2$ are all
arbitrary \e{constants} as functions of
$(\vq , \vp )$, and $\id$ is, of course, the quantum identity.  Eq.~(7)
is equivalent to Eq.~(6) with the \e{restriction} that \e{all} the
phase-space second partial derivatives of $F_1$ and $F_2$ \e{vanish}.
It is clear from the last form of the right-hand side of Eq.~(7) that
\e{by itself} Eq.~(7) yields \e{no information whatsoever} about the
quantization of classical dynamical variables.

Dirac pointed out that with the aid of Eq.~(3b) and any Taylor
series representation of $F(\vq )$, an arbitrary function of
$\vq$, his commutation postulate (e.g., Eq.~(7)) can be shown
to imply,
\re{
\{\;\Q{\vl\dt\vp}\; ,\;\Q{F(\vq)}\;\}_Q
= \Q{ \{ \vl\dt\vp, F(\vq )\} }
= -(\vl\dt\del_{\qvq}F(\qvq )).
}{8a}
Indeed, such consequences of Dirac's commutation rule can
themselves be explicitly incorporated into a further
extension of Eq.~(7), which thereupon reads, 
\re{
\{\;\Q{c + \vk\dt\vq + \vl\dt\vp}\; ,\;\Q{F(\vq) + G(\vp)}\;\}_Q
= \Q{ \{ c + \vk\dt\vq + \vl\dt\vp, F(\vq ) + G(\vp )\} }
= \vk\dt\del_{\qvp}G(\qvp ) - \vl\dt\del_{\qvq}F(\qvq ).
}{8b}
Eq.~(8b) is equivalent to Eq.~(6) with the \e{restriction} that \e{all}
the phase-space second partial derivatives of $F_1$ \e{vanish}, but
that \e{only} the $(\vq ,\vp)$--\e{mixed gradients} of $F_2$, i.e.,
those of the form $(\vl\dt\del_{\vq})(\vk\dt\del_{\vp})F_2$, need
vanish (or, alternatively, the same with $F_1$ and $F_2$ interchanged).
It is a great pity that, notwithstanding that he possessed all the
tools needed for this, Dirac apparently never wrote down and
pondered his postulate in the extended form given by
Eq.~(8b).  Had he done so, there can be no doubt that he would have
been struck by the peculiar juxtaposition of the general expression
$F(\vq) + G(\vp)$ with its mere Taylor expansion through \e{linear}
terms \e{only}, i.e., $c + \vk\dt\vq + \vl\dt\vp$.  As no quantum
physical argument requiring this decidedly unaesthetic \e{dichotomy}
suggests itself, Dirac would unquestionably have been anxious to
explore the \e{implications of removing it}---in particular whether
these implications are, \e{unlike} those of the \e{unrestricted}
Eq.~(6), \e{self-consistent}.  The resulting slightly stronger
canonical commutation rule,
\re{ \ba {c}
\{\;\Q{F_1(\vq ) + G_1(\vp )}\; ,\;\Q{F_2(\vq ) + G_2(\vp )}\;\}_Q
   =   \Q{ \{ F_1(\vq ) + G_1(\vp ), F_2(\vq ) + G_2(\vp )\} } \\
 \m{}                                                           \\
   =   \Q{\del_{\vq}F_1(\vq )\dt\del_{\vp}G_2(\vp )} -
       \Q{\del_{\vp}G_1(\vp )\dt\del_{\vq}F_2(\vq )}\: ,         \\
  \m{}
\ea }{9a}
is equivalent to Eq.~(6) with the \e{restriction} that \e{only} the
$(\vq ,\vp)$--\e{mixed gradients} of $F_1$ and $F_2$, i.e., those of
the form $(\vl\dt\del_{\vq})(\vk\dt\del_{\vp})F_i$, $i=1,2$, need
vanish.  This slightly stronger canonical commutation rule
certainly shows promise for a self-consistent
determination of the quantization of $q^2p^2$---unlike Dirac's
canonical commutation rule (given, for example, by Eq.~(8b)),
Eq.~(9a) actually implies a relation, namely,
\[  \{\,\Q{q^3}\, ,\,\Q{p^3}\,\}_Q = \Q{ \{ q^3,p^3\} }, \]
that \e{determines} the quantization of $q^2p^2$, while, \e{unlike}
the \e{unrestricted} Eq.~(6), it apparently does \e{not} permit
inference of the \e{conflicting} relation,
\[ \{\:\Q{q^2p}\: ,\:\Q{qp^2}\:\}_Q = \Q{ \{ q^2p,qp^2\} }. \]

\subsection*{Unambiguous quantization of classical dynamical variables}

More generally, the slightly stronger canonical commutation rule
of Eq.~(9a) also implies apparently unambiguous quantization
of the crucially important Fourier expansion components
$e^{i(\vk\dt\vq + \vl\dt\vp )}$ for classical dynamical variables
$F(\vq, \vp)$. It does so via the relation,
\re{
\{ e^{i\vk\dt\qvq},e^{i\vl\dt\qvp}\}_Q
   = \;\Q{ \{ e^{i\vk\dt\vq},e^{i\vl\dt\vp}\} }\;
   =   -(\vk\dt\vl)\;\Q{ e^{i(\vk\dt\vq + \vl\dt\vp )} }\: ,
}{9b}
which, together with Eq.~(5b), yields,
\re{
\Q{ e^{i(\vk\dt\vq + \vl\dt\vp )} } = (i/(\h\vk\dt\vl ))
(e^{i\vk\dt\qvq}\, e^{i\vl\dt\qvp} - e^{i\vl\dt\qvp}\, e^{i\vk\dt\qvq})\, .
}{10a}
The right hand side of Eq.~(10a) is a bit awkward in that its
limit as $\vk\rta\bv{0}$ or $\vl\rta\bv{0}$ or even $\h\rta 0$
fails to be manifestly apparent.  These obscurities can be
resolved after first formally reexpressing the commutator on
the right side as the integral of a perfect differential,
\re{
\Q{ e^{i(\vk\dt\vq + \vl\dt\vp )} } = (i/(\h\vk\dt\vl ))
\int_{0}^{1}d\a\,
\frac{d}{d\a}\lf( e^{i\a\vk\dt\qvq}\: e^{i\vl\dt\qvp}\:
 e^{i(1-\a)\vk\dt\qvq}\rt) ,
}{10b}
where, upon carrying out the differentiation with respect to $\a$,
there results,
\re{
\Q{ e^{i(\vk\dt\vq + \vl\dt\vp )} } = (-1/(\h\vk\dt\vl ))
\int_{0}^{1}d\a\, 
e^{i\a\vk\dt\qvq}\lf[\vk\dt\qvq,e^{i\vl\dt\qvp}\rt] e^{i(1-\a)\vk\dt\qvq}\, .
}{10c}
The $[\, ,]$ commutator bracket which occurs on the right hand side
of Eq.~(10c) can, using Eq.~(5b), be reexpressed as a $\{\, ,\}_Q$
quantum analog of the Poisson bracket to yield,
\re{
\Q{ e^{i(\vk\dt\vq + \vl\dt\vp )} } = (-i/(\vk\dt\vl ))
\int_{0}^{1}d\a\,
e^{i\a\vk\dt\qvq}\lf\{\vk\dt\qvq,e^{i\vl\dt\qvp}\rt\}_Q
e^{i(1-\a)\vk\dt\qvq}\, .
}{10d}
To the $\{\, ,\}_Q$ on the right hand side of Eq.~(10d) we apply
the Dirac commutation rule of Eq.~(8b) to obtain,
\re{
\Q{ e^{i(\vk\dt\vq + \vl\dt\vp )} } =
\int_{0}^{1}d\a\, e^{i\a\vk\dt\qvq}\; e^{i\vl\dt\qvp}\;
e^{i(1-\a)\vk\dt\qvq}\, .
}{10e}
It can in very similar fashion as well be shown that,
\re{
\Q{ e^{i(\vk\dt\vq + \vl\dt\vp )} } =
\int_{0}^{1}d\a\, e^{i\a\vl\dt\qvp}\; e^{i\vk\dt\qvq}\;
e^{i(1-\a)\vl\dt\qvp}\, .
}{10f}
Now since the arbitrary classical dynamical variable $F(\vq ,\vp)$ can
be \e{linearly} expanded in the Fourier components
$e^{i(\vk\dt\vq + \vl\dt\vp )}$ as,
\re{
F(\vq ,\vp ) = (2\pi )^{-2n}\int d^n\vq '\, d^n\vp '\, d^n\vk\, d^n\vl\:
F(\vq ' ,\vp ')\, e^{-i(\vk\dt\vq ' + \vl\dt\vp ')}
\, e^{i(\vk\dt\vq + \vl\dt\vp )}\, ,
}{11a}
the assumed linearity of quantization implies that its quantization
$\Q{F(\vq ,\vp )}$ is given by,
\re{
\Q{F(\vq ,\vp )} = (2\pi )^{-2n}\int d^n\vq '\, d^n\vp '\, d^n\vk\, d^n\vl\:
F(\vq ' ,\vp ')\, e^{-i(\vk\dt\vq ' + \vl\dt\vp ')}
\;\Q{ e^{i(\vk\dt\vq + \vl\dt\vp )} }\: ,
}{11b}
with the indicated quantization of $e^{i(\vk\dt\vq + \vl\dt\vp )}$
being supplied by Eq.~(10e) or (10f).

We have therefore demonstrated that the slightly strengthened canonical
commutation rule of Eq.~(9a), \e{together} with assuming \e{the linear%
ity of quantization}, implies the closed-form quantization rule of
Eq.~(11b).  Furthermore, Dirac's slightly weaker canonical commutation
rule, i.e., that of Eq.~(7) or of Eq.~(8b), obviously \e{also} follows
from the slightly stronger Eq.~(9a), and this \e{weaker} canonical com%
mutation rule of Dirac, since it \e{fails} to determine the quantiza%
tion of \e{any classical dynamical variable}, is obviously \e{consis%
tent} with the closed-form quantization rule of Eq.~(11b).

We shall now proceed to demonstrate that this \e{self-consistent} com%
bination of Dirac's canonical commutation rule and the closed-form
quantization rule of Eq.~(11b) conversely implies the \e{linearity of
quantization} and the slightly \e{strengthened} canonical commutation
rule or Eq.~(9a).  Once this demonstration is \e{completed}, we
will have shown that the \e{combination} of the slightly \e{streng%
thened} canonical commutation rule of Eq.~(9a) with the \e{linearity
of quantization} is \e{logically equivalent} to the \e{self-consistent
combination} of Dirac's canonical commutation rule with the closed-%
form quantization rule of Eq.~(11b), and therefore that the combination
of the slightly strengthened canonical commutation rule of Eq.~(9a)
with the linearity of quantization \e{is itself self-consistent}.

To begin this demonstration, we immediately note that the closed-form
quantization rule of Eq.~(11b) implies the linearity of quantization.
Furthermore, because Eq.~(10e) or (10f) implies that,
\re{
\lf.\Q{e^{i(\vk\dt\vq + \vl\dt\vp )}}\rt|_{\vl=\bv{0}} = e^{i\vk\dt\qvq}
\;\;\m{and}\;\;
\lf.\Q{e^{i(\vk\dt\vq + \vl\dt\vp )}}\rt|_{\vk=\bv{0}} = e^{i\vl\dt\qvp}\, ,
}{12a}
we have as special cases of Eq.~(11b) that,
\re{\ba{rcl}
\Q{F(\vq )} & = & (2\pi )^{-n}\int d^n\vq '\, d^n\vk\:
F(\vq ')\, e^{-i\vk\dt\vq '}\, e^{i\vk\dt\qvq}\: =\: F(\qvq )\;\;\m{and} \\
\Q{G(\vp )} & = & (2\pi )^{-n}\int d^n\vp '\, d^n\vl\:
G(\vp ')\, e^{-i\vl\dt\vp '}\, e^{i\vl\dt\qvp}\: =\: G(\qvp ). \\
            &   &
\ea}{12b}
If we further specialize $F(\vq , \vp )$ to just a constant $c$ in
the phase space argument $(\vq , \vp )$, we see that either one of
Eqs.~(12b) imply that $\q{c} = c\,\id$, where $\id$ is the quantum
identity.  Mindful of the bilinearity of $\{\, ,\}_Q$, we also see
that Eqs.~(12b) imply that,
\re{
\{ F(\qvq ),G(\qvp )\}_Q =
(2\pi )^{-2n}\int d^n\vq '\, d^n\vp '\, d^n\vk\, d^n\vl\:
F(\vq ')G(\vp ')\, e^{-i(\vk\dt\vq ' + \vl\dt\vp ')}
\,\{ e^{i\vk\dt\qvq}, e^{i\vl\dt\qvp}\}_Q.
}{12c}
The previously carried out deduction of Eq.~(10e) or (10f) from
Eq.~(10a) in concert with Dirac's canonical commutation rule of
Eq.~(8b) can readily be verified to be \e{reversible}, i.e.,
Eq.~(10a) follows from either Eq.~(10e) or (10f) together with
Dirac's canonical commutation rule of Eq.~(8b).  As Eq.~(10a) is
just Eq.~(9b) rewritten, we now put Eq.~(9b) into Eq.~(12c) to
obtain,
\re{
\{ F(\qvq ),G(\qvp )\}_Q =
(2\pi )^{-2n}\int d^n\vq '\, d^n\vp '\, d^n\vk\, d^n\vl\:
F(\vq ')G(\vp ')(-(\vk\dt\vl))\, e^{-i(\vk\dt\vq ' + \vl\dt\vp ')}
\; \Q{ e^{i(\vk\dt\vq + \vl\dt\vp )} }\: ,
}{12d}
which, via integrations by parts, can be reexpressed as,
\re{
\{ F(\qvq ),G(\qvp )\}_Q =
(2\pi )^{-2n}\int d^n\vq '\, d^n\vp '\, d^n\vk\, d^n\vl\,
(\del_{\vq '}F(\vq ')\dt\del_{\vp '}G(\vp '))\,
e^{-i(\vk\dt\vq ' + \vl\dt\vp ')}
\; \Q{ e^{i(\vk\dt\vq + \vl\dt\vp )} }\: .
}{12e}
If we now refer to Eq.~(11b), we see that Eq.~(12e) implies that,
\re{
\{ F(\qvq ),G(\qvp )\}_Q =
\Q{ \del_{\vq }F(\vq )\dt\del_{\vp }G(\vp ) } =
\Q{ \{ F(\vq ), G(\vp )\} }\: ,
}{12f}
which is the key part of the slightly stronger canonical commutation
rule given by Eq.~(9a).  The \e{remainder} of Eq.~(9a) follows
from the properties of the Poisson bracket, the linearity of quanti%
zation, and the relations,
\de{
\{ F_1(\qvq ),F_2(\qvq )\}_Q = 0\;\;\m{and}\;\;
\{ G_1(\qvp ),G_2(\qvp )\}_Q = 0,
}
which are an obvious consequence of Dirac's canonical commutation
rule given by Eq.~(8b).

As noted in the paragraphs following Eq.~(11b), this completes
the demonstration of the \e{self-consistency} of the slightly
strengthened canonical commutation rule of Eq.~(9a) in concert
with the linearity of quantization, which, of course, together
are \e{logically equivalent} to the closed-form quantization
rule of Eq.~(11b) taken in concert with Dirac's slightly weaker
canonical commutation rule of Eq.~(7) or (8b).

Thus the slightly stronger canonical commutation rule of Eq.~(9a)
\e{navigates to perfection} the tight, perilous channel between
the Scylla of self-inconsistently \e{overdetermining} quantization
by placing \e{insufficient restriction} on the \e{classical corres%
pondence} embodied by Eq.~(6) and the Charybdis of ambiguously
\e{underdetermining} quantization by anxiously \e{overrestricting}
the classical correspondence which flows from Eq.~(6), as Dirac did
when he limited the inputs of Eq.~(6) to \e{only the Cartesian com%
ponents} of the phase space vector $(\vq ,\vp)$.

In retrospect, it seems truly astonishing that Dirac himself, in
the course of his long theoretical physics career, did not event%
ually hit upon the modest and completely natural upgrade of his
inadequate canonical quantization postulate to the vastly more
satisfactory Eq.~(9a)---this is surely a cautionary real life
lesson in the fact that even so penetrating and innovative a mind
as was Dirac's can still become snagged in a conceptual side stream.
Probably the first to discover Eq.~(9a) was E.H. Kerner~\ct{Kr}%
, who realized that it confirmed a groundbreaking clarification
of Hamiltonian path integral quantization which he had, together
with W.G. Sutcliffe, pointed out~\ct{K-S}%
.  Unfortunately, Kerner, who was apparently a very sensitive
individual, never published this discovery, possibly because his
earlier path integral work with Sutcliffe had met with strong---%
albeit entirely misconceived---opposition~\ct{Co}
(to which he as well declined to riposte, so as not, in his words,
``to pick a fight''~\ct{Kr}%
).  Formal further development and elaboration of the path inte%
gral quantization insight of Kerner and Sutcliffe is addressed in
another manuscript, but it is interesting to note that the Hamil%
tonian path integral naturally incorporates a maximally strong
type of the classical correspondence: the single \e{most} important
path which enters into the path sum is the one of stationary phase,
and that path is also \e{always} the \e{classical} phase-space path
which obeys Hamilton's equations of motion.  It is perhaps not sur%
prising, then, that the results which flow from such a principle of
maximum classical correspondence dovetail with those that result
from requiring the maximum correspondence between quantum and clas%
sical Poisson brackets which is self-consistent.  Now because the
physically correct Hamiltonian path integral in configuration space
directly yields the quantization of its input classical Hamiltonian
in \e{configuration representation}, we shall here work out the
quantization of an arbitrary classical dynamical variable in that
particular representation.  As a preliminary step, we work out
the configuration space matrix elements of the quantization of
the classical Fourier component $e^{i(\vk\dt\vq + \vl\dt\vp )}$
that is given by Eq.~(10e),
\re{
\langle\vq_2 |\;\Q{ e^{i(\vk\dt\vq + \vl\dt\vp )} }\; |\vq_1\rangle =
\int_{0}^{1}d\a\: e^{i\vk\dt (\a\vq_2 + (1-\a )\vq_1 )}
\; \delta^{(n)}(\vq_2 - \vq_1 + \h\vl ),
}{13a}
which, together with Eq.~(11b), yields these same matrix elements
of the quantization of an arbitrary classical dynamical variable
$F(\vq ,\vp )$,
\re{
\langle\vq_2 |\;\Q{F(\vq ,\vp )}\; |\vq_1\rangle =
(2\pi\h )^{-n}\int d^n\vp
\int_{0}^{1}d\a\, F(\a\vq_2 + (1-\a )\vq_1 , \vp )
\: e^{i\vp\dt(\vq_2 - \vq_1 )/\h}\, .
}{13b}
Though it is not well-known, there is also a Hamiltonian path
integral in \e{momentum space}, which directly yields the
quantization of its input classical Hamiltonian in
\e{momentum representation}.  For this reason, we use Eqs.~%
(10f) and (11b) to likewise work out the momentum space
matrix elements of the quantization of our arbitrary
classical dynamical variable $F(\vq ,\vp )$,
\re{
\langle\vp_2 |\;\Q{F(\vq ,\vp )}\; |\vp_1\rangle =
(2\pi\h )^{-n}\int d^n\vq
\int_{0}^{1}d\a\, F(\vq ,\a\vp_2 + (1-\a )\vp_1 )
\: e^{-i\vq\dt(\vp_2 - \vp_1 )/\h}\, .
}{13c}
The quantization formulas we have derived in Eqs.~(13b) and
(13c) on the basis of the strengthened canonical commutation
postulate of Eq.~(9a) arise \e{entirely naturally} as well from
the Hamiltonian path integral \e{provided} that care is taken
to \e{ensure} that the paths summed over \e{all actually adhere}
to the two imposed endpoint constraints $\vq (t_1) = \vq_1$ and
$\vq (t_2) = \vq_2$ (or, for the momentum-space path integral,
$\vp (t_1) = \vp_1$ and $\vp (t_2) = \vp_2$).  It was
R. P. Feynman himself who introduced a very widely used approx%
imating sequence of path sets which each have paths that 
\e{inadvertently deviate} from these physically crucial con%
straints by \e{arbitrarily large} amounts, no matter \e{how far}
one goes along that path set sequence~\ct{Fe}%
.  The work of Kerner and Sutcliffe ensures that approximating
path sets for the Hamiltonian path integral are free of such
egregiously unphysical elements~\ct{K-S}%
.

\subsection*{The generalized Ehrenfest theorem}

Because Dirac's inadequate canonical commutation postulate prevents
determination of the quantization of the vast majority of classi%
cal dynamical variables, it precludes even \e{investigation} of the
very interesting issue of whether the Ehrenfest theorem relation of
the mean dynamical behavior of a quantized system to Hamilton's
classical equations of motion for that system is a \e{universal}
fact.  The slightly stronger canonical commutation rule of
Eq.~(9a), however, \e{always} determines $\Q{F(\vq ,\vp )}$ through
Eq.~(11b) in conjunction Eq.~(10a), (10e), or (10f).  From any of
these latter three equations, it is readily seen that,
\re{
\{ \qvq ,\;\Q{e^{i(\vk\dt\vq + \vl\dt\vp )}}\;\}_Q =
 i\vl\,\Q{e^{i(\vk\dt\vq + \vl\dt\vp )}} \;\;\m{and}\;\;
\{ \qvp ,\;\Q{e^{i(\vk\dt\vq + \vl\dt\vp )}}\;\}_Q = 
 -i\vk\,\Q{e^{i(\vk\dt\vq + \vl\dt\vp )}}\: ,
}{14a}
which, in conjunction with Eq.~(11b), implies that,
\re{
\{ \qvq ,\;\Q{F(\vq ,\vp )}\;\}_Q = 
 \Q{\del_{\vp}F(\vq ,\vp )} \;\;\m{and}\;\;
\{ \qvp ,\;\Q{F(\vq ,\vp )}\;\}_Q = 
 -\,\Q{\del_{\vq}F(\vq ,\vp )}\: .
}{14b}
Eq.~(14b), together with the quantum equation of motion (1b)
for arbitrary quantized dynamical variables, implies that,
\re{
d\qvq /dt =
 \Q{\del_{\vp}H(\vq ,\vp )} \;\;\m{and}\;\;
d\qvp /dt =
 -\,\Q{\del_{\vq}H(\vq ,\vp )}\: .
}{15a}
Taking arbitrary expectation values of both sides of the two
equations in (15a) shows that the Ehrenfest theorem indeed
applies \e{universally} to Hamiltonian dynamical systems,
\re{
\lf\langle d\qvq /dt \rt\rangle =
\lf\langle \Q{\del_{\vp}H(\vq ,\vp )} \rt\rangle \;\;\m{and}\;\;
\lf\langle d\qvp /dt \rt\rangle =
-\lf\langle \Q{\del_{\vq}H(\vq ,\vp )} \rt\rangle ,
}{15b}
which provides an elegant counterpoint to the Correspondence
Principle.

\subsection*{Noninvertibility of quantization}

It is also quite interesting to note
that the quantization given by Eq.~(11b) and either Eq.~(10e) or
(10f), or, equivalently, by Eq.~(13b) or Eq.~(13c), has
the property that every quantized dynamical variable has \e{more}
than one classical precursor (indeed it has an uncountable infinity
of them), which implies that quantization is \e{not} an invertible
procedure.  We demonstrate this by exhibiting an uncountable number
of nontrivial classical precursors to the identically \e{zero}
quantized dynamical variable, from which the more general result
follows because of the \e{linearity} of quantization.  To find this
plethora of nontrivial classical precursors to the quantized zero,
we merely need to look among the members of the now familiar class
of nontrivial dynamical variables of the form $e^{i(\vk\dt\vq +
\vl\dt\vp )}$, where the vector pair $(\vk ,\vl )$ runs over $R^{2n}$.
The members of this class are obviously quantized by Eq.~(10e) or
Eq.~(10f).  The exponentiated operators in the right hand sides of
these equations can be combined by the repeated application of the
simplest special case of the Campbell-Baker-Hausdorff formula, namely,
\re{
e^{i\q{F}}\: e^{i\q{G}} = e^{-\hf c}\: e^{i(\q{F} + \q{G})}\;\;
\m{when $[\q{F},\q{G}] = c\,\id$, where c is a constant.}
}{16}
After consolidation of the exponentiated operators in Eq.~(10e) or
(10f), the integration over the variable $\a $ can be straightfor%
wardly carried out to yield,
\re{
\Q{ e^{i(\vk\dt\vq + \vl\dt\vp )} } =
\frac{\sin (\hf\h (\vk\dt\vl ))}{(\hf\h (\vk\dt\vl ))}
\; e^{i(\vk\dt\qvq + \vl\dt\qvp )}\, .
}{17a}
Alternatively, the integration over the variable $\a $ in Eq.~(13a)
can also be straightforwardly carried out, and then followed by the
application of a trigonometric identity to yield,
\re{
\langle\vq_2 |\;\Q{ e^{i(\vk\dt\vq + \vl\dt\vp )} }\; |\vq_1\rangle =
\frac{\sin (\hf\h (\vk\dt\vl ))}{(\hf\h (\vk\dt\vl ))}
\; e^{i(\vk\dt (\vq_1 + \vq_2 ))/2}
\; \delta^{(n)}(\vq_2 - \vq_1 + \h\vl ).
}{17b}
It is clear from both Eq.~(17a) and (17b) that,
\re{
\Q{ e^{i(\vk\dt\vq + \vl\dt\vp )} } = 0\;\;
\m{when $(\vk\dt\vl ) = 2n\pi /\h$, where $n$ is any \e{nonzero}
signed integer.}
}{17c}
Therefore the quantized zero has an uncountable number of nontrivial
classical dynamical precursors, and the same obviously also holds
for \e{any} quantized dynamical variable.  One might wonder how this
state of affairs can be compatible with the Correspondence Princi%
ple---the answer involves the highly nonuniform asymptotic ``conver%
gence'' which is so typical of the correspondence ``limit''.  One
sees from Eq.~(17c) and the Schwarz inequality that those classical
dynamical variables $e^{i(\vk\dt\vq + \vl\dt\vp )}$ whose quantiza%
tions vanish have $\vk $ and $\vl $ that satisfy $|\vk ||\vl |\geq
2\pi/\h $.  Thus, in the limit that $\h\rta 0$, we will have, for
these classical dynamical precursors of quantized zero, that $|\vk |
\rta\infty $ or $|\vl |\rta\infty $ or both, which causes these clas%
sical dynamical variables $e^{i(\vk\dt\vq + \vl\dt\vp )}$ to oscillate
arbitrarily rapidly.  It is in this highly nonuniform asymptotic sense
that these nontrivial classical precursors of \e{quantized} zero
indeed also ``wash out'' to zero in the correspondence ``limit''
$\h\rta 0$.  In the \e{quantum} world, however, all dynamical variables
formally correspond to an infinite number of classical precursors,
and thus have a kind of ``many classical potentialities'' aura which
is eminently compatible with at least the \e{spirit} of the Uncertainty
Principle.

\subsection*{The quantization supposition of Born and Jordan}

The quantization rule that is given by Eq.~(11b) and either
Eq.~(10e) or (10f) was historically first presented by Born and Jordan
in the paper which sets out their intriguing \e{variational} development
of quantum mechanics~\ct{B-J}%
.  That development absorbs Heisenberg's equation of motion into a
variational principle which involves a trace rather than an integral,
and has as its consequence commutation rules which not only predated
those of Dirac, but are also, in principal, stronger than his---in%
deed Eq.~(14b) can be regarded as the central \e{consequence} of the
systematic Born-Jordan development of quantum mechanics.  It so hap%
pens, however, that the commutation rules of Eq.~(14b) are \e{also}
compatible with quantization rules which \e{differ} from the ``Born%
-Jordan'' one embodied by Eq.~(11b) and either Eq.~(10e) or
(10f)---they are, for example, compatible with Weyl's quite differ%
ent quantization rule, which we shall discuss next.  Thus the dis%
covery of ``Born-Jordan'' quantization by those authors must be re%
garded as serendipitously premature---the central tenets of the
quantum mechanics they developed, while entirely \e{compatible} with
this quantization, do \e{not} uniquely \e{imply} it.  (Born and Jor%
dan, however, were apparently unaware of that fact.)

\subsection*{Contrasts with Weyl's maximally austere quantization}

The mathematician Weyl, who took an interest in the nascent quantum
mechanics, realized that Dirac's canonical commutation rules
failed to pin down the quantization of a general classical
dynamical variable, and decided to try his hand at rectifying that
situation.  Weyl immediately grasped the essence of Eq.~(11b); i.e.,
that with the assumption that quantization is a linear process, one
need only determine the quantizations of the classical dynamical
Fourier component functions $e^{i(\vk\dt\vq + \vl\dt\vp )}$~\ct{We}%
.  Mathematician that he was, Weyl applied \e{no} further
\e{physics-related} considerations \e{whatsoever} to this problem,
but in archtypical fashion aimed to construct the \e{formally} most
straightforward, spartan, and elegant quantization of these Fourier
component functions possible.  Not surprisingly, given this thrust,
Weyl arrived at the quantization postulate,
\re{
\Q{ e^{i(\vk\dt\vq + \vl\dt\vp )} } = e^{i(\vk\dt\qvq + \vl\dt\qvp )}\, ,
}{18a}
to be used in conjunction with Eq.~(11b).  Application of Eq.~(16)
shows that Eq.~(18a) can also be written as,
\re{
\Q{ e^{i(\vk\dt\vq + \vl\dt\vp )} } =
e^{i\vk\dt\qvq /2}\; e^{i\vl\dt\qvp}\; e^{i\vk\dt\qvq /2} =
e^{i\vl\dt\qvp /2}\; e^{i\vk\dt\qvq}\; e^{i\vl\dt\qvp /2}\, ,
}{18b}
which permits direct comparison of Weyl's quantization rule
with the Born-Jordan quantization Eq.~(10e) or (10f).  It is
thus seen that Weyl's choice is the \e{single} most \e{symmetrical}
operator ordering of a quite general class, whereas Born-Jordan
quantization is the \e{equally weighted average} of \e{all} the
operator orderings of that class.  Notwithstanding that Weyl
undoubtedly achieved his goal of spartan elegance, it becomes
clear that the character of the quantum physics can involve yet
profounder themes---Born-Jordan quantization's even-handed embrace
of \e{all} orderings of the class in question seems to echo the path
integral's sum over \e{all} applicable paths.

In the manner of Eq.~(13b) for Born-Jordan quantization of the
classical dynamical variable $F(\vq ,\vp )$ in configuration
representation, one has the following Weyl quantization in
configuration representation of $F(\vq ,\vp )$,
\re{
\langle\vq_2 |\;\Q{F(\vq ,\vp )}\; |\vq_1\rangle =
(2\pi\h )^{-n}\int d^n\vp\:
F(\hf (\vq_1 + \vq_2 ), \vp )
\: e^{i\vp\dt(\vq_2 - \vq_1 )/\h}\, .
}{19a}
Noting that Weyl's Eq.~(18a) exhibits none of the mapping into
quantized zero of classical dynamical Fourier components which is
so apparent for its Born-Jordan counterpart of Eq.~(17a), we need
to seriously entertain the possibility that Weyl's quantization is
one-to-one and \e{invertible}.  This indeed becomes rather apparent
upon reexpressing Eq.~(19a) in terms of the variables
$\vq_- = (\vq_2 - \vq_1 )$ and $\vq_+ = \hf (\vq_1 + \vq_2 )$,
\re{
\langle\vq_+ + \hf\vq_- |\;\Q{F(\vq ,\vp )}\; |\vq_+ - \hf\vq_- \rangle =
(2\pi\h )^{-n}\int d^n\vp\:
F(\vq_+ , \vp )
\: e^{i\vp\dt\vq_- /\h}\, ,
}{19b}
which reveals a straightforward Fourier transformation of $F(\vq ,\vp )$
in just its second independent variable $\vp $.  This transformation is
easily inverted to recover the classical $F(\vq ,\vp )$ itself,
\re{
F(\vq , \vp ) =
\int d^n\vq_-\,
\langle\vq + \hf\vq_- |\;\Q{F(\vq ,\vp )}\; |\vq - \hf\vq_- \rangle
\: e^{- i\vp\dt\vq_- /\h}\, .
}{19c}
Eqs.~(19b) and (19c) not only show that Weyl's quantization is one-%
to-one and invertible, but (19c) also shows that inverse to be given
by the well-known ``classical Wigner representation'' for quantized
dynamical variables.  The \e{unique} emergence of the Wigner repre%
sentation at this juncture is something less than a resounding
\e{physical} endorsement of Weyl's quantization---it is, for example,
well-known that there exist positive definite quantum operators which
have classical Wigner representations that attain negative values
over very significant regions of phase space.  Even the straightfor%
ward apparent elegance of one-to-one invertible quantization
\e{itself} frays a little around the edges on closer physical scrutiny.
If one takes the \e{unique} classical precursor of each quantized dy%
namical variable seriously, one will need to wrestle philosophically
with the \e{determinism} of the consequent well-defined classical
``shadow world''---note that because the quantized \e{operators} evolve
deterministically under the influence of their Heisenberg equations of
motion, their \e{unique} classical ``shadows'' will do the same.  On
balance, it would seem wisest to studiously ignore the antics of Weyl's
unique classical precursors. This, however, raises the question of the
scientific appropriateness of entertaining at all a theoretical
hypothesis that gives rise to a prominent mathematical feature
(such as invertibility by Wigner representation), whose most
\e{obvious} physical interpretations conflict with \e{other}
tenets of known physical theory---particularly when there exist
\e{alternative} theoretical hypotheses (e.g., Born-Jordan quant%
ization) which simply do \e{not} give rise to that mathematical
feature and which are, in \e{addition}, far more extensively based
on \e{physics-related} arguments.

Finally, Weyl's quantization \e{also} turns out to be the result of a
sequence of approximating path sets (due to Feynman~\ct{Fe}) for the
Hamiltonian path integral which \e{inadvertently} has the property
that \e{all} of those sets contain paths which have \e{arbitrarily
large deviations} from the two (physically crucial!) imposed endpoint
constraints.  Born-Jordan quantization, in utter contrast, is the
result of a sequence of approximating path sets (due to Kerner and
Sutcliffe~\ct{K-S}) for the Hamiltonian path integral which has the
property that \e{each and every path} scrupulously \e{conforms} to
those two imposed endpoint constraints.  Ironically, routine
\e{unsuspecting} use of the endpoint constraint \e{breaching}
Feynman sequence of approximating path sets for Hamiltonian path
integrals has long been a mainstay of support for a widely shared
assumption that Weyl quantization is probably physically
\e{correct}~\ct{Fe}, \e{notwithstanding} the well-known woes of the
thereupon \e{unavoidable} Wigner representation, which Feynman in
particular was even moved to desperately try to make sense of through
an attempt to ``interpret'' negative probabilities!


\begin{thebibliography}{9}
\bibitem{Di}
P. A. M. Dirac,
Proc.\ Roy.\ Soc.\ (London) \textbf{A109},
642 (1925);
\textbf{A110},
561 (1926);
\e{The Principles of Quantum Mechanics}
(Oxford University Press, London, 1947).
\bibitem{Gr}
H. J. Groenewold,
Physica \textbf{12},
405 (1946).
\bibitem{Kr}
E. H. Kerner,
private conversation.
\bibitem{K-S}
E. H. Kerner and W. G. Sutcliffe,
J.\ Math.\ Phys.\ \textbf{11},
391 (1970).
\bibitem{Co}
L. Cohen,
J.\ Math.\ Phys.\ \textbf{11},
3296 (1970).
\bibitem{Fe}
R. P. Feynman,
Phys.\ Rev.\ \textbf{84},
108 (1951).
\bibitem{B-J}
M. Born and P. Jordan,
Z.\ Physik \textbf{34},
858 (1925).
\bibitem{We}
H. Weyl,
Z.\ Physik \textbf{46},
1 (1927).
\end{thebibliography}
\end{document}